# An Agent-Based Modeling for Pandemic Influenza in Egypt

Khaled M. Khalil, M. Abdel-Aziz, Taymour T. Nazmy, Abdel-Badeeh M. Salem
Faculty of Computer and Information Science Ain shams University Cairo, Egypt
kmkmohamed@gmail.com, mhaziz67@gmail.com, ntaymoor@yahoo.com,
absalem@asunet.shams.edu.eg

**Abstract**

*Pandemic influenza has great potential to cause large and rapid increases in deaths and serious illness. The objective of this paper is to develop an agent-based model to simulate the spread of pandemic influenza (novel H1N1) in Egypt. The proposed multi-agent model is based on the modeling of individuals' interactions in a space-time context. The proposed model involves different types of parameters such as: social agent attributes, distribution of Egypt population, and patterns of agents' interactions. Analysis of modeling results leads to understanding the characteristics of the modeled pandemic, transmission patterns, and the conditions under which an outbreak might occur. In addition, the proposed model is used to measure the effectiveness of different control strategies to intervene the pandemic spread.*

## 1. Introduction

The first major pandemic influenza H1N1 is recorded in 1918-1919, which killed 20-40 million people and is thought to be one of the most deadly pandemics in human history. In 1957, a H2N2 virus originated in China, quickly spread throughout the world and caused 1-4 million deaths world wide. In 1968, an H3N2 virus emerged in Hong Kong for which the fatalities were 1-4 million [16]. In recent years, novel H1N1 influenza has appeared. Novel H1N1 influenza is a swine-origin flue and is often called swine flue by the public media. The novel H1N1 outbreak began in Mexico, with evidence that there had been an ongoing epidemic for months before it was officially recognized as such. It is not known when the epidemic will occur or how sever it will be. Such an outbreak would cause a large number of people to fall ill and possibly die.

In the absence of reliable pandemic detection systems, computer models and systems have become an important information tools for both policy-makers and the general public [15]. Computer models can help in providing a global insight of the infectious disease outbreaks' behavior by analyzing the spread of infectious diseases in a given population, with varied geographic and demographic features [12]. Computer models promise an improvement in representing and understanding the complex social structure as well as the heterogeneous patterns in the contact networks of real-world populations determining the transmission dynamics [4]. One of the most recent approaches of such sophisticated modeling is agent-based modeling [3]. Agent-based modeling of pandemics recreates the entire populations and their dynamics through incorporating social structures, heterogeneous connectivity patterns, and meta-population grouping at the scale of the single individual [3].

In this paper we propose a stochastic multi-agent model to mimic the daily person-to-person contact of people in a large scale community affected by a pandemic influenza in Egypt. The proposed model is used to: (i) assess the understanding of transmission dynamics of pandemic influenza, (ii) assess the potential range of consequences of pandemic influenza in Egypt, and (iii) assess the effectiveness of different pandemic control strategies on the spread of the pandemic. We adopt disease parameters and the recommended control strategies from WHO [16]. While, we use Egypt census data of 2006 [7] to create the population structure, and the distribution of social agent attributes. Section 2 reviews different epidemiological modeling approaches: mathematical modeling, cellular automata based modeling, and agent based modeling. While, section 3 reviews related multi-agent models in literature. Section 4 discusses the proposed model, and section 5 validates the proposed model. Section 6 discusses the pandemic control strategies and their effect on the spread of the pandemic. Section 7 presents the modeling experiments and analysis of results, and then we conclude in Section 8.

## 2. Epidemiological modeling approaches

The search for an understanding of the behavior of infectious diseases spread has resulted in several attempts to model and predict the pattern of many different communicable diseases through a population [5]. The earliest account was carried out in 1927 by Kermack and McKendrick [9]. Kermack and McKendrick created a mathematical model named SIR (Susceptible-Infectious-Recovered) based on ordinary differential equations. Kermack and McKendrick started with the assumption that all members of the community are initially equally susceptible to the disease, and that a complete immunity is conferred after the infection. The population is divided into three distinct classes (see figure 1): the susceptible (S) healthy individuals who can catch the disease; the infectious (I) those who have the disease and can transmit it; and the recovered (R) individuals who have had the disease and are now immune to the infection (or removed from further propagation of the disease by some other means).

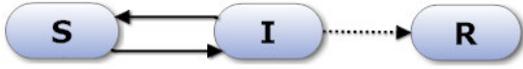

Figure 1. SIR (Susceptible–Infectious–Recovered) Model

Let $S(t), I(t)$, and $R(t)$ be the number of susceptible, infected and recovered individuals, respectively, at time *t*, and *N* is the size of the fixed population, so we have:

$$N = S(t) + I(t) + R(t) \qquad (1)$$

Upon contact with an infected a susceptible individual contracts the disease with probability $\beta$, at which time he immediately becomes infected and infectious (no incubation period); infectious recover at an individual rate $\gamma$ per unit time. Based on mentioned assumptions; Kermack and McKendrick derived the classic epidemic SIR model as follows:

$$\begin{aligned} \frac{dS}{dt} &= -\beta SI \\ \frac{dI}{dt} &= \beta SI - \gamma I \qquad (2) \\ \frac{dR}{dt} &= \gamma I \end{aligned}$$

From equations (1) and (2), we found that SIR model is deterministic and doesn't study the nature of population vital dynamics (handling newborns and deaths). Following Kermack and McKendrick, other physicians contributed to modern mathematical epidemiology; extending SIR model with more classes and supporting vital dynamics such as: SEIR (Susceptible–Exposed–Infectious–Recovered), and MSEIR (Immunized–Susceptible–Exposed–Infectious–Recovered) models [9]. However, mathematical models had not taken into account spatial and temporal factors such as variable population structure, and dynamics of daily individuals' interactions which drive more realistic modeling results [1].

The second type of developed models is cellular automata based models, which incorporate spatial parameters to better reflect the heterogeneous environment found in nature [13]. Cellular automata based models are an alternative to using deterministic differential equations, which use a two-dimensional cellular automaton to model location specific characteristics of the susceptible population together with stochastic parameters which captures the probabilistic nature of disease transmission [2]. Typically a cellular automaton consists of a graph where each node is a cell. This graph is usually in the form of a two-dimensional lattice whose cells evolve according to a global update function applied uniformly over all the cells [13]. Cell state takes one of the SIR model states, and is calculated based on cell present state and the states of the cells in its interaction neighborhood. As the Cellular automata based model evolves, cells states will determine how the overall behavior of a complex system [2]. However, cellular automata based models neglect the social behavior and dynamics interactions among individuals in the modeled community. Therefore, cellular automata gave way to a new approach; Agent-based models.

Agent-based models (ABM) are similar to cellular automata based models, but leverage extra tracking of the effect of the social interactions of individual entities [1]. Agent-based model consists of a population of agents, an environment, and set of rules managing agents' behavior [12]. Each agent has two components: a state and a step function. Agent state describes every agent attributes values at the current state. The step function creates a new state (usually stochastically) representing the agent attributes at the next time step. The great benefit of agent-based models that these models allow epidemiological researchers to do a preliminary "what-if" analysis with the purpose of assessing systems' behavior under various conditions and evaluating which alternative control strategies to adopt in order to fight epidemics [12].

## 3. Multi-agent related models

Related agent-based models are Perez-Dragicevi model, BIOWar, and EpiSims. Perez-Dragicevi [12] had developed a multi-agent model to simulate the spread of a communicable disease in an urban environment using measles outbreak in an urban environment as a case study. The model uses SEIR (Susceptible–Exposed–Infectious–Recovered) model and makes use of census data of

Canada. The goal of this model is to depict the disease progression based on individuals' interactions through calculation of ratios of susceptible/infected in specific time frames and urban environments. BIOWar [10] is a computer model that combines computational models of social networks, communication media, and disease transmission with demographically resolved agent models, urban spatial models, weather models, and a diagnostic error model to produce a single integrated model of the impact of a bioterrorist attack on an urban area. BIOWar models the population of individual agents as they go about their lives. BIOWar allows the study of various attacks and containment policies as revealed through indicators such as medical phone calls, insurance claims, death rates, over-the counter pharmacy purchases, and hospital visit rates, among others. EpiSims [8] is an agent-based model, which combines realistic estimates of population mobility, based on census and land-use data of USA, with configurable parameters for modeling the progress of a disease. EpiSims involves a way to generate synthetic realistic social contact networks in a large urban region.

However, the proposed agent-based model will differ from the above models for: (i) usage of census data of Egypt, (ii) proposed extension to SIR model, (iii) and studying the effect of different control strategies on the spread of the disease. This study is considered very important which incorporate Egypt population structure in modeling process. We have adopted Egypt census data of 2006 for creating realistic social contact networks such as home, work and school networks. While, the proposed extension to SIR model encompasses new classes; modeling the real pandemic behavior and control states such as (in contact, quarantined, not quarantined, dead, and immunized). In addition, we involve the study of the effect of different control strategies on the spread of the pandemic influenza. We plan in future work to integrate the proposed model with different simulation tools and models such as weather models, transportation models, and decision support models to build a complete system for pandemic management in Egypt.

## 4. Proposed model

In what follows, we propose an extension to SIR model. Then we propose the multi-agent model based on the proposed extension of SIR model states. Finally, we validate the proposed multi-agent model by aligning with the classical SIR model.

### 4.1. Proposed extension to SIR model

We propose an extension to SIR model by adding extra classes to represent more realistic agent states (see figure 2). In addition, we adopt stochastic approach to traverse among agent states using normal distribution. Agents are grouped based on the proposed extension to SIR model into nine classes. The first class is the (S) Susceptible agents, who are not in contact with infectious agents and are subject to be infected. At the start of the modeling, all agents fall in the (S) Susceptible class. The second class is the (C) in Contact agents, who are in direct contact with other infectious agents. The third class is the (E) Exposed agents, who are infected agents during the incubation time (latent) of the disease. The fourth class is the (I) Infectious agents, who are contagious. The fifth class is the (Q) Quarantined agents, who are infected agents quarantined by the health care authorities. The sixth class is the (NQ) Not Quarantined agents, who are infected agent but not quarantined. The seventh class is the (D) Dead agents. The eighth class is the (R) Recovered agents. The ninth class is the (M) Immunized agents, who are immunized against the disease infection.

Figure 3 presents flow chart which explains in details the sequence of the state chart of the proposed extension to SIR model. All population members are born susceptible then may contact contagious agents (move into in-contact class). In-contact agents may acquire the infection (move into the exposed class) based on given distribution. Exposed agents remain non-contagious for given latent time. At the end of the latent time, agents will become contagious (move into the infectious class). Infected agents may ask for doctor help and thus become quarantined by health care authorities (move into quarantined class), or ignore disease symptoms (move to non-quarantined class) based on given distribution. Non-quarantined agents are the main source of disease in this model. Non-quarantined agents may ask for doctor help and thus become quarantined, or die (move to dead class) based on given distribution. Quarantined agents may response to disease drugs and become recovered (move to recovered class) or die (move to dead class) based on given distribution. Recovered agents may become immunized (move to immunized class) based on given distribution, or become susceptible again.

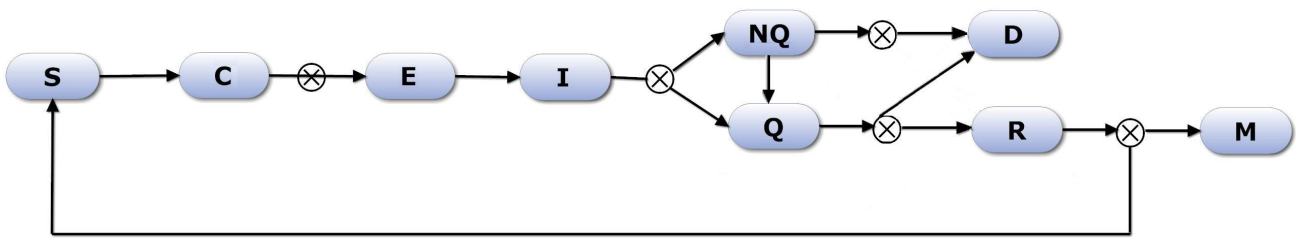

Figure 2. State chart of proposed extension to SIR model. (S) Susceptible, (C) in Contact, (E) Exposed, (I) Infectious, (Q) Quarantined, (NQ) Not Quarantined, (D) Dead, (R) Recovered, and (M) Immunized.

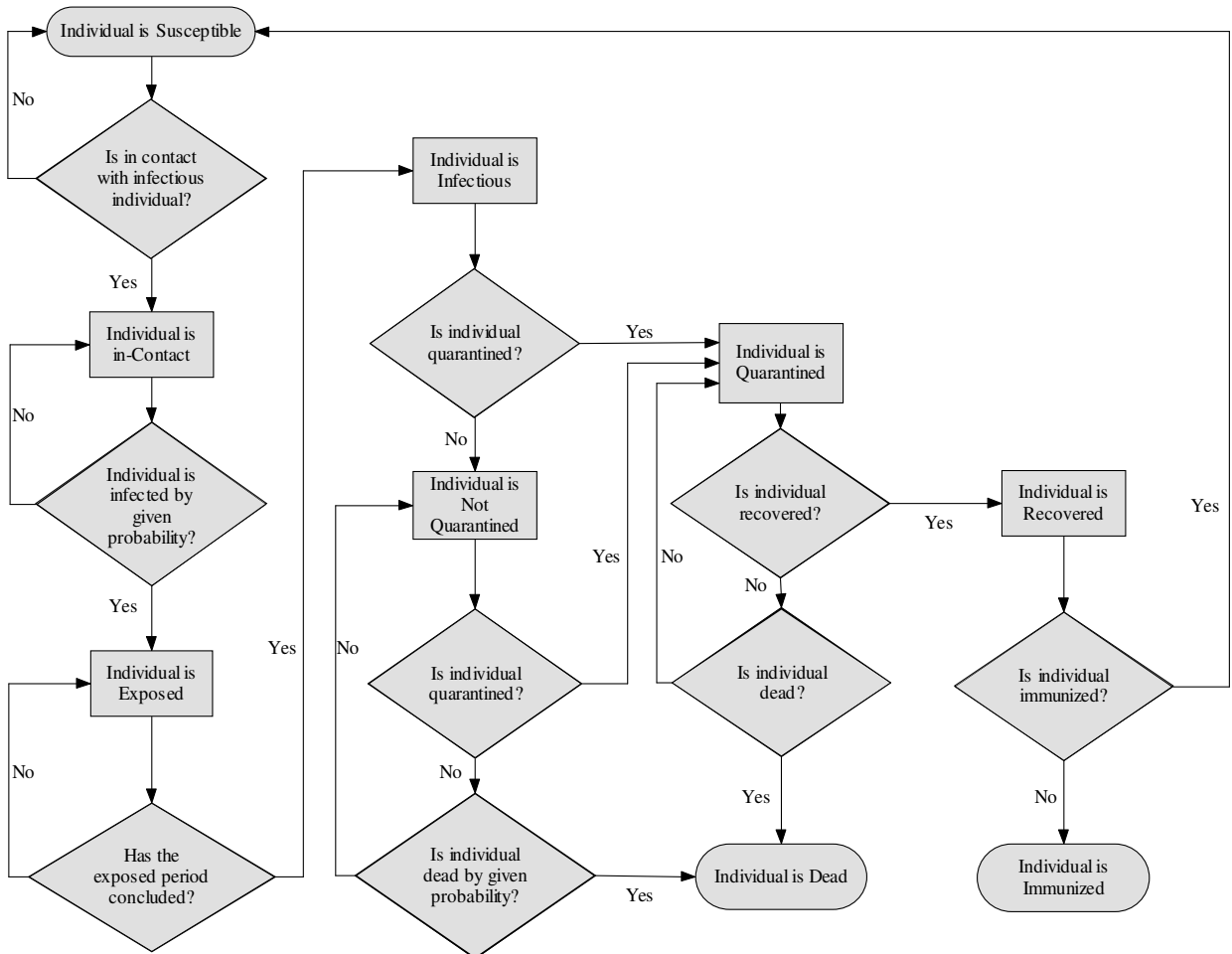

Figure 3. Flow chart of the proposed extension to SIR model.

## 4.2. Proposed multi-agent model

The proposed multi-agent based model attempts to realistically represent the behavior of individuals' daily activities, and the natural biological process of the pandemic influenza spread among individuals as a result of individuals' interactions. Proposed agent-based model involves (i) population agents, (ii) agents' rules which govern the behavior of the agents, (iii) and the infection transmission patterns following the proposed extension to SIR model. Agents represent human population, in which each agent is involved in a sequence of daily basis activities according to the agent social type. These daily activities allows agents to interact themselves in groups or even travel and join other groups. The daily activities of working, travelling, and public gathering are modeled, while agents' states are calculated on discrete time steps during agent life time.

Proposed multi-agent model has several parameters such as: simulation parameters, disease model parameters, agents' attributes, and population distribution based on census data. First: simulation parameters which include (i) number of days to be simulated, (ii) random seed for the gaussian random number generator, (iii) population size, (iv) and initial agents. Second: disease model parameters which include: (i) incubation time which is the average time of infected agent before being contagious, (ii) percentage of recovered infected agents after treatment, (iii) percentage of immunized agents after the recovery, (iv) percentage of dead agents, (v) average minimum and maximum time required to recover infected agent, and finally (vi) percentage of quarantined infected agents (see Table 1 for parameters default values). Third: agent attributes which are crucial for describing the nature of the pandemic and control the behavior of agent among time and space. Agent attributes includes: (i) health state (based on proposed extension to SIR Model states), (ii) social activities level (High, Moderate, Low), (iii) daily movement, (iv) spatial location, (v) infection time, (vi) social type (SPOUSE, PARENT, SIBLING, CHILD, OTHERFAMILY, COWORKER, GROUPMEMBER, NEIGHBOR, FRIEND, ADVISOR, SCHOOLMATE, OTHER), and (vii) agent social networks. Social activities level controls the number of daily contacts of the agent, which proportionally affect the number of in-contact agents interacting with the infected agent. Increasing number of in-contact agents adds more chance to the reproduction number to increase which means epidemic outbreak [4]. Reproduction number (R) can be defined as the average number of secondary agents infected by a primary agent case [16]. We have distributed number of daily infected cases based on the social activities level as following: Low: 2 agents, Moderate: 3 agents, and High 4 agents. Social type distribution is based on Egypt Census data of 2006 [7]. Egypt census data has classified population into five classes (see Table 2). The distribution of social types based on census data leads to different social network structure. All involved distributions are assumed to be Gaussian distribution with mean of 0 and standard deviation of 1. Distributions are subject to be replaced in future work according to the availability of more information about Egypt population.

At the beginning of the system startup, configurations are loaded and user is required to set up initial agents. The simulation runs in a loop for a pre-specified number of days. When the simulation starts, the first step is to create the initial agents instances. Next, agents start practicing their natural daily activities. During each day, every member of the agent community moves around, and communicates with their social network agents or with public agents. See figure 4 for the spatial representation of agents moving and contacting each other, and their health states represented with different icons. Daily moved distance by agents and the number of contacted agents are randomly determined by each agent attributes. During the day activities, simulation keeps track of the social networks of each agent which can be at work, home, or school.

Table 1: Default values of model parameters

| Parameter | Default Value |
| --- | --- |
| Number of simulated days | 50 |
| Random Seed | 0 |
| Population Size | 72798031[*] |
| Agents | Initial agents |
| Incubation Time (latent) | 2 days[**] |
| Percentage of immunized agents | 0.95[**] |
| Percentage of recovered agents | 0.9[**] |
| Percentage of dead agents | 0.14[**] |
| Min. time to be recovered | 5 day's[*] |
| Max. time to be recovered | 14 days[**] |
| Percentage of quarantined agents | 0.1 |

[*] Egypt census data of 2006 [7].
[**] WHO [16] values for novel H1N1 pandemic.

Selection of contact agents is random and most likely results in contacting new agents who are created at runtime. Newly created agents are initially susceptible, and are placed randomly across the landscape. Agent social types are drawn from normal distribution based on the population structure of census data of Egypt 2006. Agent health state and disease clock are changed according to the proposed extension to SIR model. Infected agent will affect all agents in his social networks to be exposed. Thus, probability of agent to be infected increases according to the number of infected agents in his social networks.

## 5. Proposed model validation

It is often very difficult to validate epidemiological simulation models due to the lack of reliable field data, and the lack of real geographical location of the individual cases occurred. We have to validate the proposed multi-agent model against other available models that have been validated such as SIR model [14]. SIR Model has a long history and has proved to be a plausible model for real epidemics. The proposed model should be aligned with the SIR model at least for some simplified scenarios. In order to align the proposed model to SIR model, we have evaluated SIR model using Mathematica [11] based on given parameters (basic reproduction number $R_0 = 3$, duration of Infection = 9.5, initially immunized = 0, initially infected = 0.01% - see Figure 5) and we have evaluated the same model parameters in the proposed multi-agent model (see Figure 6). Two graphs are not a perfect match, but the proposed multi-agent model graph match the general behavior of SIR model graph. Two graphs differ by the magnitude and the smoothness of the curves. The source of difference of curves behavior is confined in the following factors: (i) the heterogeneous structure of the population, (ii) different reproduction numbers which are calculated for each agent independently, (iii) and the usage of random variable for infection time instead of deterministic values in SIR model.

Table 2. Age distribution of Egypt

| Age (years) | Percentage | Possible social types |
| --- | --- | --- |
| < 4 | 10.60 % | SIBLING, CHILD, OTHER |
| 05–14 | 21.10 % | SIBLING, CHILD, OTHERFAMILY, COWORKER, GROUPMEMBER, NEIGHBOR, FRIEND, SCHOOLMATE, OTHER |
| 15–44 | 49.85 % | SPOUSE, PARENT, SIBLING, OTHERFAMILY, COWORKER, GROUPMEMBER, NEIGHBOR, FRIEND, ADVISOR, SCHOOLMATE, OTHER |
| 45–59 | 12.36 % | SPOUSE, PARENT, SIBLING, OTHERFAMILY, COWORKER, GROUPMEMBER, NEIGHBOR, FRIEND, ADVISOR, OTHER |
| > 59 | 6.08 % | SPOUSE, PARENT, SIBLING, OTHERFAMILY, GROUPMEMBER, NEIGHBOR, FRIEND, OTHER |

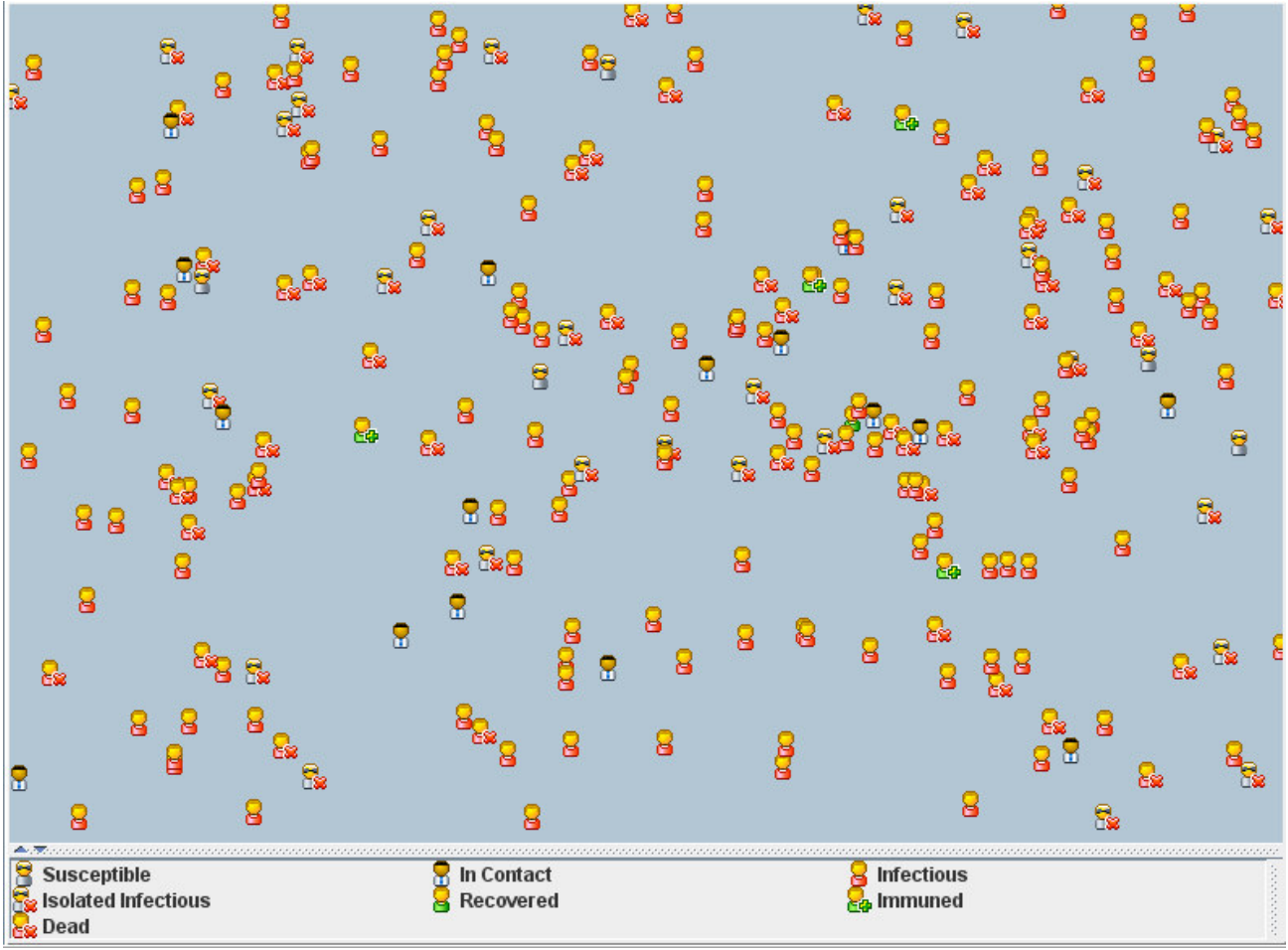

Figure 4. Snapshot displays agents with different health states moving and contacting each others.

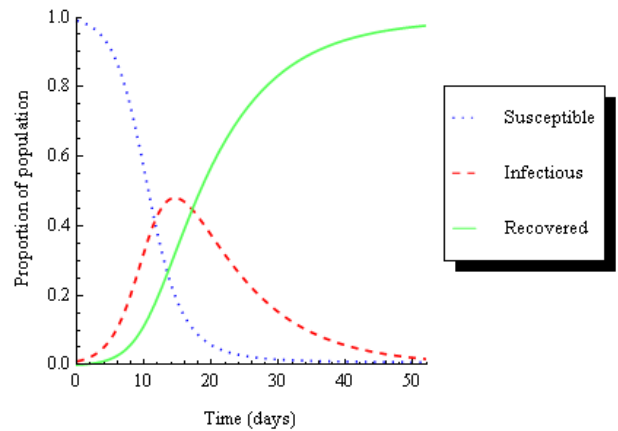

Figure 5. Mathematica SIR Model. Parameters: $R_0=3$, Duration of Infection=9.5, initially immunized = 0, initially infected = 0.01%.

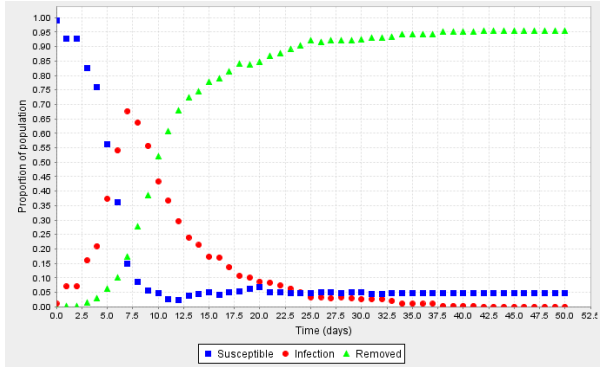

Figure 6. Proposed multi-agent model.
Parameters: 3 infected agents – Population Size: 300 agents – Duration: 50 days.

## 6. Pandemic control strategies

Control strategies are useful for the development of an action plan to control disease outbreak. Controlling outbreak is related to the peak of infectious and the time required reaching the peak. Time required reaching the peak is helpful for giving time for different control strategies to be effective. Without a properly planned strategy, the pandemic chaos might be disastrous causing large-scale fatalities and substantial economic damage. A proven control strategy would incorporate increasing awareness of population, vaccination, social distancing, and quarantining decisions [16] [6]. The proposed multi-agent model permits injection of control strategies to study different scenarios for controlling the outbreak. User can determine control strategies and the coverage percentage applied to the population. Control strategies will affect the agent health states, and the agent daily activities. Increasing awareness will increase the number of doctors' visit and the number of quarantined infected agents. Vaccination moves susceptible and in-contact agents to be immunized. Social distancing and quarantining reduce the number of possible contacts among individual agents. In experiments we will run different scenarios of applied control strategies.

## 7. Experiments and analysis of results

The proposed multi-agent model was programmed using Java programming language and run on AMD Athlon 64 X2 Dual 2.01 GHz processor with 1GB memory. To demonstrate the model behavior, we have run five scenarios of pandemic influenza in a closed population of 1000 agents, and initially three infected agents. Each run takes about eight minutes to be completed. Simulated scenarios are: (i) population with no deployed control strategies, (ii) population with deployed 50% of increasing awareness control strategy, (iii) population with deployed 50% of vaccination control strategy, (iv) population with deployed 50% of social distancing control strategy, and (v) population with deployed 50% of quarantining control strategy. We display susceptible, infectious and removed curves of each scenario to be compared with each other. In the first scenario with no deployed control strategies, we have found that epidemic has a steep infection curve which reaches its peak (608 infected agents – 60.8% of the population is infected) on day 10 (see figure 7). At the end of the simulation, we have analyzed the distribution of health states among social types. Mortality rate is high among schoolmates, neighbors, and advisors. Numbers of immunized agents are close for schoolmates, and parents, while equals zero with child agents (see figure 8).

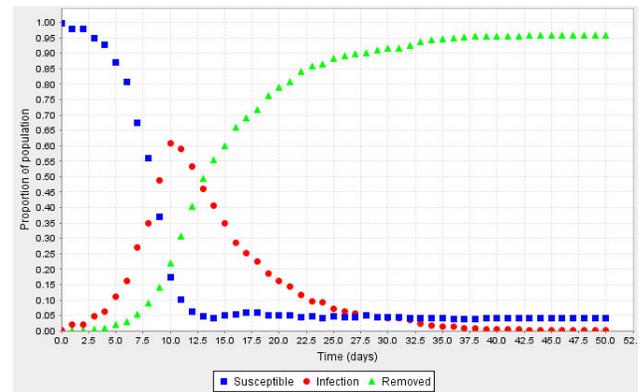

Figure 7. Scenario 1: pandemic peak is at day 10.

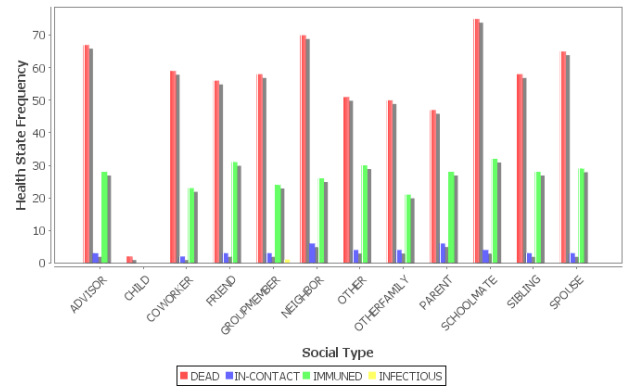

Figure 8. Scenario 1: distribution of health states to social types.

We have a pandemic peak at day 10. Thus, we choose to apply control strategies from day 8 to day 12 in the rest of the scenarios to study the effect of the control strategy on the pandemic peak. In the second scenario, we found that the number of infected agents is reduced during the deployment period of increase awareness control strategy to reach it is minimum value of 67 infected agents (6.7 % of the population) at day 12 (see figure 9.a). This is because agents are asking for doctor help when they have influenza symptoms.

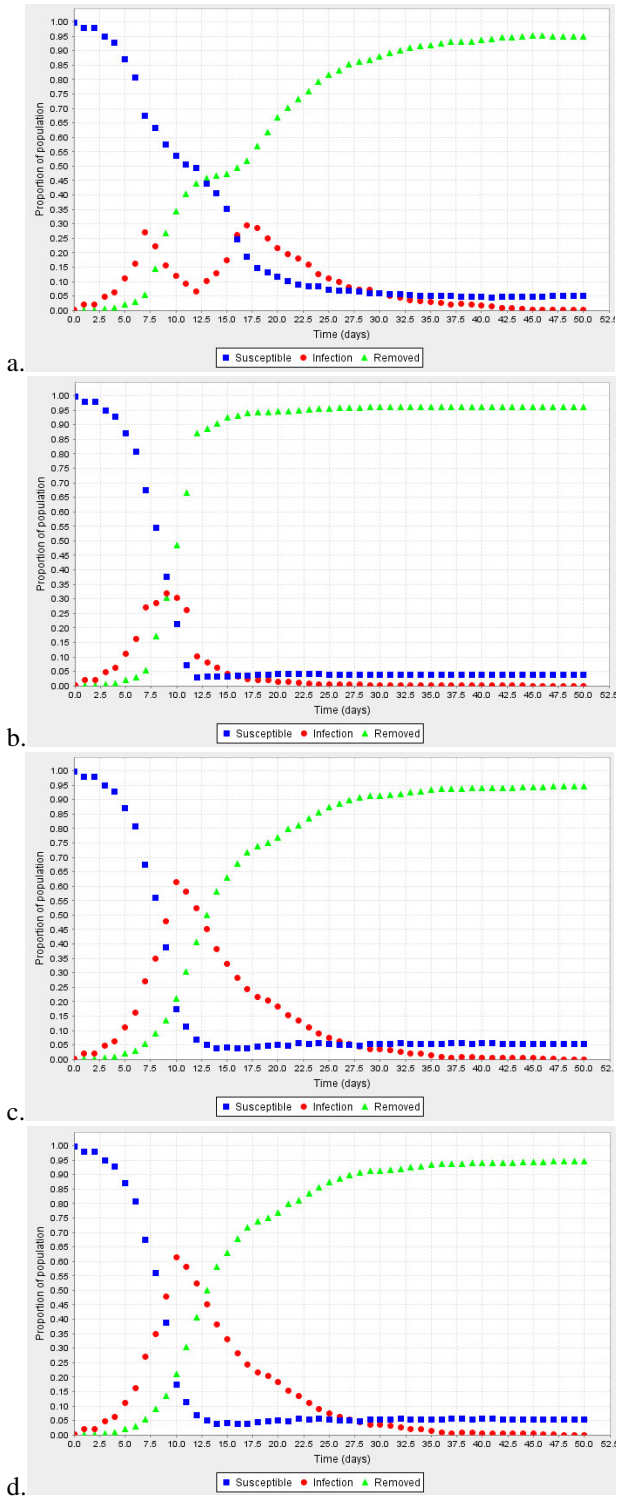

a.
b.
c.
d.

Figure 9. a: Scenario 2, b: Scenario 3, c: Scenario 4, d: Scenario 5.

In the third scenario, we found that the number of infected agents is reduced to 320 agents (32% of the population) at day 9 (see figure 9.b). In the fourth and the fifth scenarios, we found that the number of infected agents is increased to 614 (61.4% of the population) at day 10 (see figure 9.c - d). This means that there are control strategies which not affect the pandemic spread when applied on given outbreak duration such as: social distancing and quarantining through the outbreak peak. Finally, we conclude that the aggregate attack rate is exponentially increasing without any deployed control strategies. Attack rate is controlled by different factors such as: the daily travelling distance of agents, and the percentage of vaccinated agents. As a result, proper combination of deployed control strategies can be effective to decrease the pandemic damage.

## 8. Conclusion

The field of computational epidemiology has arisen as a new branch of epidemiology to understand epidemic transmission patterns, and to help in planning precautionary measure. For this reason a spatially explicit agent-based epidemiologic model of pandemic contagious disease is proposed in this paper. The methodology for this paper involves the development of a multi-agent model of pandemic influenza in Egypt. The proposed model simulates stochastic propagation of pandemic influenza outbreaks, and the impact of the decisions made by the healthcare authorities in population with millions of agents. We have proposed extension to SIR model, in which we have investigated the agent attributes. The model can be easily customized to study the pandemic spread of any other communicable disease by simply adjusting the model parameters. We have simulated the spread of novel H1N1 pandemic in Egypt. Experiments are run in a closed population of 1000 agents, and initially 3 infected agents. Modeled novel H1N1 reaches infection peak (608 agents) with in 10 days without deployment of control strategies. Number of dead agents reaches its peak at the end of the simulation with mortality of 658 dead agents. Deployment of proper combination of control strategies can limit the pandemic chaos and reduce the fatalities and substantial economic damage. Further work on proposed model includes: agents with additional attributes that allow a better realistic model (e.g., ages, gender, etc), as well as finding optimal combination of control strategies to manage the pandemic outbreak waves.